\documentclass[a4paper,11pt]{article} 
\usepackage{amssymb}  
\usepackage{amsthm}
\usepackage{amsmath}
 \usepackage{graphicx}
  \usepackage{subcaption}
  \usepackage{relsize}
  \usepackage{float}

  \newtheorem{theorem}{Theorem}[section]
\newtheorem{corollary}{Corollary}[theorem]

\theoremstyle{definition}
\newtheorem{definition}{Definition}[section]
\newtheorem{exam}{Example}[section]

\usepackage{etoolbox}

\makeatletter
\patchcmd{\maketitle}{\@fnsymbol}{\@alph}{}{}  
\makeatother

\title{Diffusion of a particle via a stochastic process on the Pascal'e pyramid.}
\author{
 Pei-wen Kao\thanks{Email:peggy.kao.06@gmail.com}
  }
\date{}

\begin{document}

\maketitle 



\begin{abstract}
In this note, we construct a $3$-dimensional generalisation of the Pascal's triangle that we named Pascal's cube, as it has the construction of a cube with entries given by extended binomial coefficients ${}^cC^{a}_{b}$. The Pascal's cube is equivalent to the well-studied Pascal's pyramid, with the advantage that the Pascal's cube can be mapped onto the Cartesian plane for easier computation.  

We define a stochastic process using extended binomial coefficients on the Pascal's pyramid representing the dispersion of a free particle. With some constrains, we showed that this stochastic process satisfies the heat equation.  

\end{abstract}
%
%


\section{Introduction}

A stochastic process is a rule for assigning every experimental outcome a function $P(x,t)$, which gives rise to a family of time functions depending on the position $x$ \cite{stochastic processes}. Brownian motion and random walks are examples of a stochastic process.  

It has been of great interest to connect quantum mechanics to a stochastic theory of Physics. For instance,  \cite{Millard1988} is an attempt to link the Schr\"{o}dinger equation to a Markov process, which is a stochastic process in $6$ dimension hypothesised for the motion of a particle.      

Since the Pascal's triangle is well-known to present the possibility of random walks on a $2$-dimensional coordinate \cite{William1968}, it is intriguing to find a connection between the Pascal's triangle and quantum mechanics. For instance, \cite{FFS2014} introduced a generalisation of the classical Pascal's triangle and presented a link between a stochastic process and quantum mechanics. 

In this paper, we present a link between a stochastic process via the Pascal's pyramid, a higher dimensional generalisation of the Pascal's triangle and heat equation. Since the heat equation is equivalent to the Schr\"{o}dinger equation with a wick rotation in time, there could be a possible connection between stochastic processes on the Pascal's pyramid and quantum mechanics.   

This paper is organised as follows. In section 2, we construct Pascal's cube, a variation of the Pascal's pyramid with entries denoted by extended binomial coefficients ${}^cC^{a}_{b}$ which can be computed using binomial coefficients. In section 3 we maps the Pascal's cube to the Pascal's pyramid. In section 4, a stochastic process $P(x,y,t)$ is defined as the probability distribution on the Pascal's pyramid, describing the dispersion of a particle with time. In section 5, we construct a stochastic process $P(x,t)$ on the Pascal's pyramid satisfying the heat equation.    

\section{Pascal's cube and extended binomial coefficients}

In this section, we construct Pascal's cube and introduce extended binomial coefficients.

\subsection{Construction of the Pascal's cube}

\begin{figure}[H]
\begin{subfigure}{.5\textwidth}
  \centering
  \includegraphics[width=.9\linewidth]{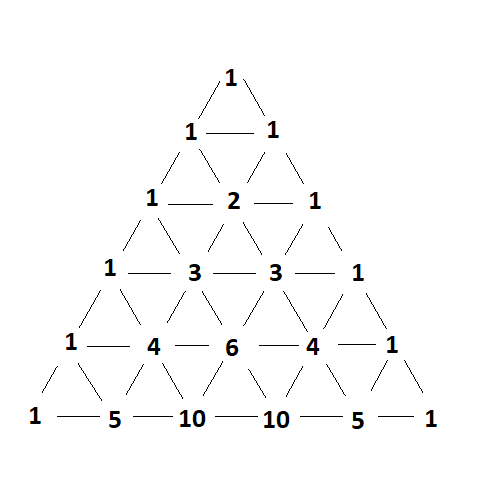}
  \caption{The first $6$ rows of Pascal's triangle.}
  \label{fig:triangle}
\end{subfigure}
\begin{subfigure}{.5\textwidth}
  \centering
  \includegraphics[width=.9\linewidth]{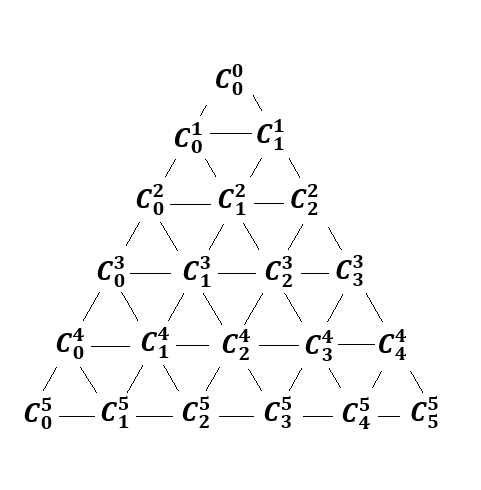}
  \caption{Pascal's triangle represented by binomial coefficients.}
  \label{fig:triangle cab}
\end{subfigure}
\caption{Pascal's triangle}
\label{Pascal triangle}
\end{figure}

Pascal's triangle, named after the French Mathematician and Philosopher Blaise Pascal \cite{Smith1929}, is a triangular array of the binomial coefficient arranged by adding the two closest numbers above. Fig \ref{fig:triangle} displays the first $6$ rows of the Pascal's triangle. 

Pascal's triangle arises in combinatorics as each entry in the Pascal's triangle can be represented by a binomial coefficient $C_{k}^n$ \cite{Smith1929}, where $C_{k}^n$ is defined by
\begin{displaymath}
C_{k}^n=\frac{n!}{k!(n - k)!}.
\end{displaymath} 
Thus the Pascal's triangle can be represented by binomial coefficients as shown in Fig \ref{fig:triangle cab}. 

It follows from the definition of the Pascal's triangle that
 \begin{equation}
   \left\{ \begin{array}{ll}
 C_{0}^n & = C^{n}_{n} =  1, \\
 C_{a}^n & = C_{a-1}^{n-1} + C_{a}^{n-1}, 
 \end{array} \right. \label{pascal triangle}
 \end{equation}
 where $n \ge a+1, a \ge 1$, $n, a \in \mathbb{Z}$.

\begin{figure}[H]
\begin{subfigure}{.5\textwidth}
  \centering
  \includegraphics[width=1\linewidth]{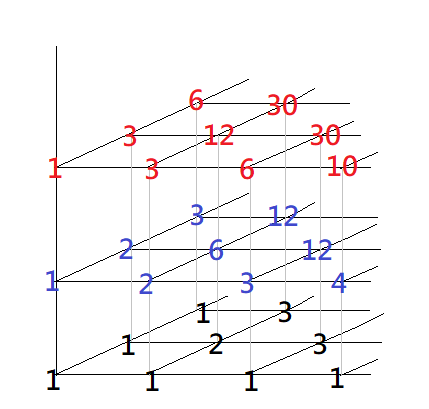}
  \caption{Pascal's cube represented by numbers.}
  \label{fig:3D number}
\end{subfigure}
\begin{subfigure}{.5\textwidth}
  \centering
  \includegraphics[width=1\linewidth]{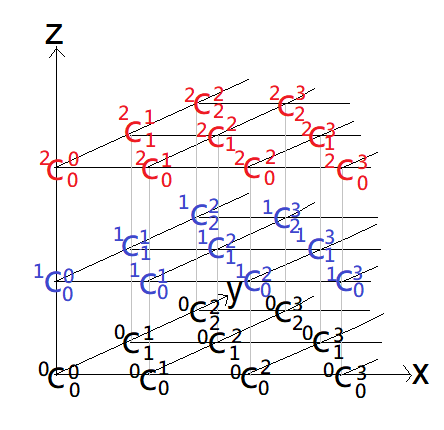}
  \caption{Pascal's cube represented by extended binomial coefficients.}
  \label{fig:3D number com}
\end{subfigure}
\caption{Pascal's cube, a three dimensional generalisation of the Pascal's triangle.}
\label{fig:Pascal cube}
\end{figure}

We then generalise the Pascal's triangle to $3$-dimensions by adding one dimension perpendicular to the Pascal's triangle, and each entry is arranged by adding the three closest prior numbers. Let us name the $3$-dimensional generalisation of the Pascal's triangle a Pascal's cube, as it takes the shape of a cube. The first three layers of the Pascal's cube is presented in Fig \ref{fig:3D number}. 

As shown in Fig \ref{fig:3D number com}, we define elements in the Pascal's cube by ${}^{c}C_{b}^{a}$, where $a,b,c \in \mathbb{Z}_{+}$. ${}^{c}C_{b}^{a}$ is called an extended binomial coefficient. The base layer of the Pascal's cube is the original Pascal's triangle, such that ${}^0C^{a}_{b} \equiv C^{a}_{b}$.

\begin{definition}[Extended binomial coefficient]
As a generalisation of \eqref{pascal triangle}, elements in the Pascal's cube are denoted by ${}^{c}C_{b}^{a}$ and are defined by the following relationships:
\begin{equation}
 \left\{ \begin{array}{ll}
{}^cC^{0}_{0} &= 1,\\
{}^cC^{a}_{0} &= {}^cC^{a-1}_{0}+{}^{c-1}C^{a}_{0},\\
{}^cC^{a}_{a} &= {}^cC^{a-1}_{a-1}+{}^{c-1}C^{a}_{a},\\
{}^cC^{a}_{b} &= {}^cC^{a-1}_{b-1}+{}^cC^{a-1}_{b}+{}^{c-1}C^{a}_{b},
 \end{array} \right. \label{3dC}
\end{equation}
where $a>b>0$.
\end{definition}

Similar to the binomial coefficients, extended binomial coefficients satisfy the following symmetry
\begin{eqnarray*}
{}^cC^{0}_{0} &=& C^{0}_{0},\\
{}^{c}C^{a}_{b} &=& {}^{c}C^{a}_{a-b}.
\end{eqnarray*}

We can put a Cartesian plane on the Pascal's cube as follows:
 \begin{corollary}
Consider the Cartesian coordinate system on the Pascal's cube as shown in Fig. \ref{fig:3D number com}. The extended binomial coefficient at a given coordinate $(x,y,z)$ is given by ${}^{z}C^{x+y}_{y}$. 
 \end{corollary} 

\subsection{Extended binomial coefficients}

Extended binomial coefficients are entries of the Pascal's cube satisfying Eqn. \ref{3dC}. In this session, we demonstrate that these coefficients can be computed using the usual binomial coefficients. 

For the $c$-th layer, here is a general rule for the extended binomial coefficient:

\begin{theorem}\label{extended binomial th}

\begin{eqnarray*}
{}^{c}C^{a}_{b} &=& \sum_{n=0}^{a}\sum_{m=0}^{b}C^{n}_{m} . {}^{c-1}C^{a-n}_{b-m}\\
&=& \prod_{i=0}^{c} \left( \sum_{n_{i}}^{a}\sum_{m_{i}}^{b} C_{m_{i}}^{n_{i}}\right) \times C^{a-n}_{b-m}, 
\end{eqnarray*}
where $n=\sum_{i=0}^{c}n_{i}$, and $m=\sum_{i=0}^{c}m_{i}$. 

With the following restrictions imposing on $n_{i}$ and $m_{i}$:
\begin{eqnarray*}
&& n_{i} \ge m_{i} \ge 0, \\
&&  a-(n_{1}+n_{2}+n_{3}+\cdots +n_{c}) \ge b-(m_{1}+m_{2}+m_{3}+\cdots+m_{c}),\\
&& 0 \le \sum_{i=1}^{c} n_{i} \le a, \text{and   } 0 \le \sum_{i=1}^{c} m_{i} \le b. 
\end{eqnarray*}
\end{theorem}
\begin{proof}
\textit{Base case:} 
\begin{eqnarray*}
{}^{c}C^{0}_{0} &=& C^{0}_{0} . C^{0}_{0} = 1,\\
{}^{c}C^{a}_{0} &=& \sum_{n=0}^{a} C^{n}_{0} . {}^{c-1}C^{a-n}_{0} = \sum_{n=0}^{a} {}^{c-1}C_{0}^{a-n}={}^cC^{a-1}_{0}+{}^{c-1}C^{a}_{0},\\
{}^{c}C^{a}_{a} &=& {}^{c}C^{a}_{0}. 
\end{eqnarray*}

\textit{Inductive step:}
Assuming 
\begin{equation*}
{}^{c}C^{a}_{b} = \sum_{n=0}^{a}\sum_{m=0}^{b}C^{n}_{m} . {}^{c-1}C^{a-n}_{b-m}.
\end{equation*}
Then
\begin{eqnarray*}
{}^{c}C^{a+1}_{b} &=& {}^{c}C^{a}_{b-1}+{}^{c}C^{a}_{b}+{}^{c-1}C^{a+1}_{b} \\
&=& \sum_{n=0}^{a}\sum_{m=0}^{b-1} C^{n}_{m} . {}^{c-1} C^{a-n}_{b-1-m} 
+ \sum_{n=0}^{a}\sum_{m=0}^{b} C^{n}_{m} . {}^{c-1}C^{a-n}_{b-m}
+ \sum_{n=0}^{a+1}\sum_{m=0}^{b} C^{n}_{m} . {}^{c-2} C^{a-n}_{b-m}\\
&=& \sum_{n=0}^{a+1}\sum_{m=0}^{b}C^{n}_{m}\left( {}^{z-1} C^{a-n}_{b-1-m} + {}^{c-1}C^{a-n}_{b-m} + {}^{c-2} C^{a-n}_{b-m} \right)\\
&=&  \sum_{n=0}^{a+1}\sum_{m=0}^{b} C^{n}_{m} . {}^{c-1} C^{a+1-n}_{b-m}.
\end{eqnarray*}

Thus by mathematical induction, theorem \ref{extended binomial th} holds for all $c \ge 0$.
\end{proof}

Following is an example to demonstrate this rule.

\begin{exam}
\begin{eqnarray}
{}^{2}C_{2}^{3} &=& \sum_{n_{1}=0}^{3}\sum_{n_{2}=0}^{3-n_{1}}\sum_{m_{1}=0}^{2}\sum_{m_{2}=0}^{2-m_{1}} C_{m_{1}}^{n_{1}} C_{m_{2}}^{n_{2}} C_{2-m_{1}-m_{2}}^{3-n_{1}-n_{2}} \nonumber\\
&=& C_{0}^{0} \left( C_{0}^{0}C_{2}^{3} + C_{0}^{1}C_{2}^{2} + C_{1}^{1}C^{2}_{1} + C^{2}_{1} C^{1}_{1} + C^{2}_{2}C^{1}_{0} + C^{3}_{2}C^{0}_{0}
\right) \nonumber\\
&+& C^{1}_{0} \left( C^{0}_{0}C^{2}_{2} + C^{1}_{1}C^{1}_{1}+C^{2}_{2}C^{0}_{0} \right) + C^{1}_{1} \left( C^{0}_{0} C^{2}_{1} + C^{1}_{0} C^{1}_{1} + C^{1}_{1}C^{1}_{0} +C^{2}_{1} C^{0}_{0} \right) \nonumber\\
&+& C^{2}_{1} \left( C^{0}_{0} C^{1}_{1} + C^{1}_{1} C^{0}_{0} \right) 
+ C^{2}_{2} \left( C^{0}_{0}C^{1}_{0} +C^{1}_{0} C^{0}_{0} \right) + C^{3}_{2} C^{0}_{0} C^{0}_{0}\nonumber\\
&=& 30. \nonumber
\end{eqnarray}
\end{exam}

\section{Mapping Pascal's cube to Pascal's pyramid}

\begin{figure}[H]
\begin{subfigure}{.5\textwidth}
  \centering
  \includegraphics[width=0.9\linewidth]{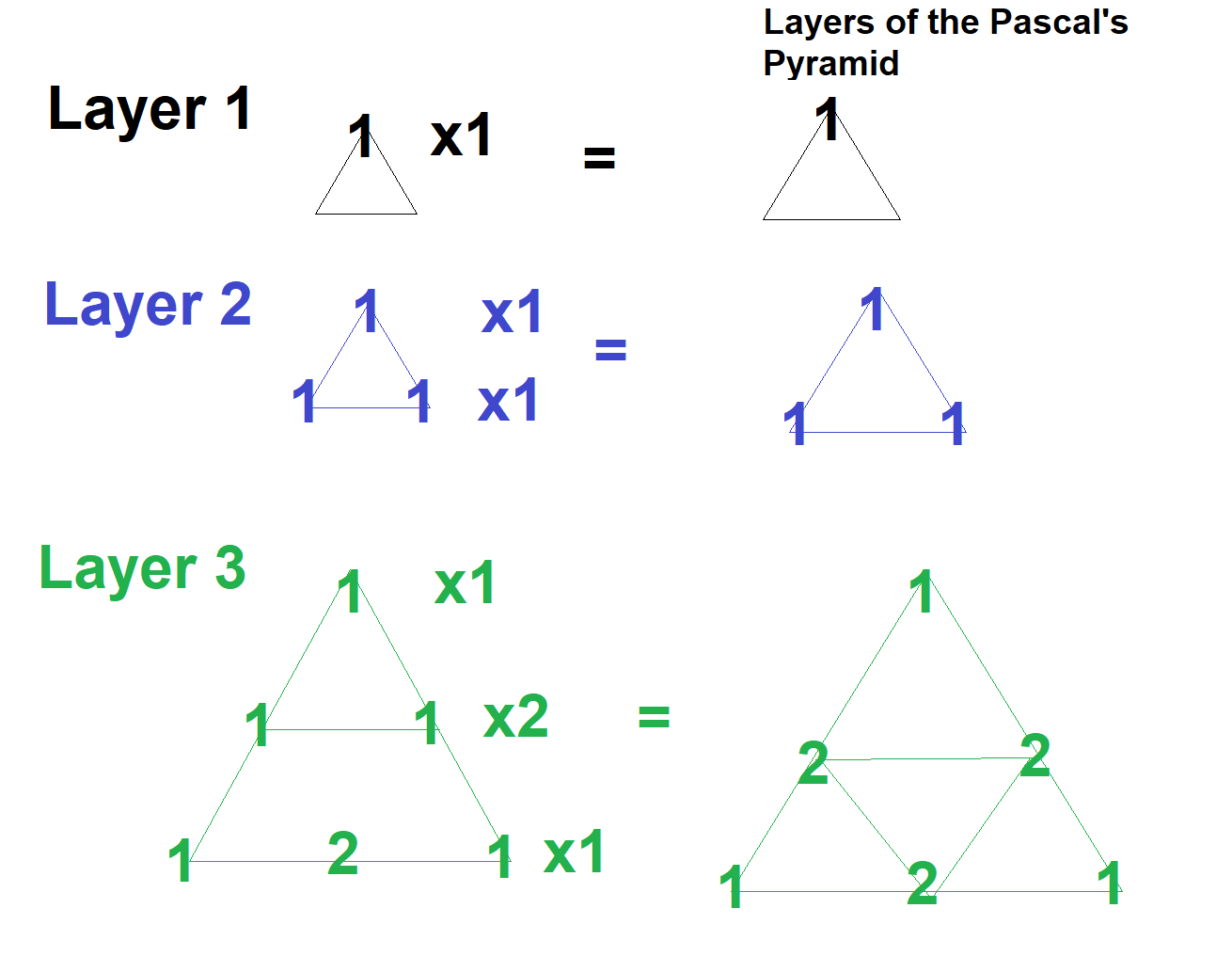}
  \caption{First $3$-layers.}
\end{subfigure}
\begin{subfigure}{.5\textwidth}
  \centering
  \includegraphics[width=1.0\linewidth]{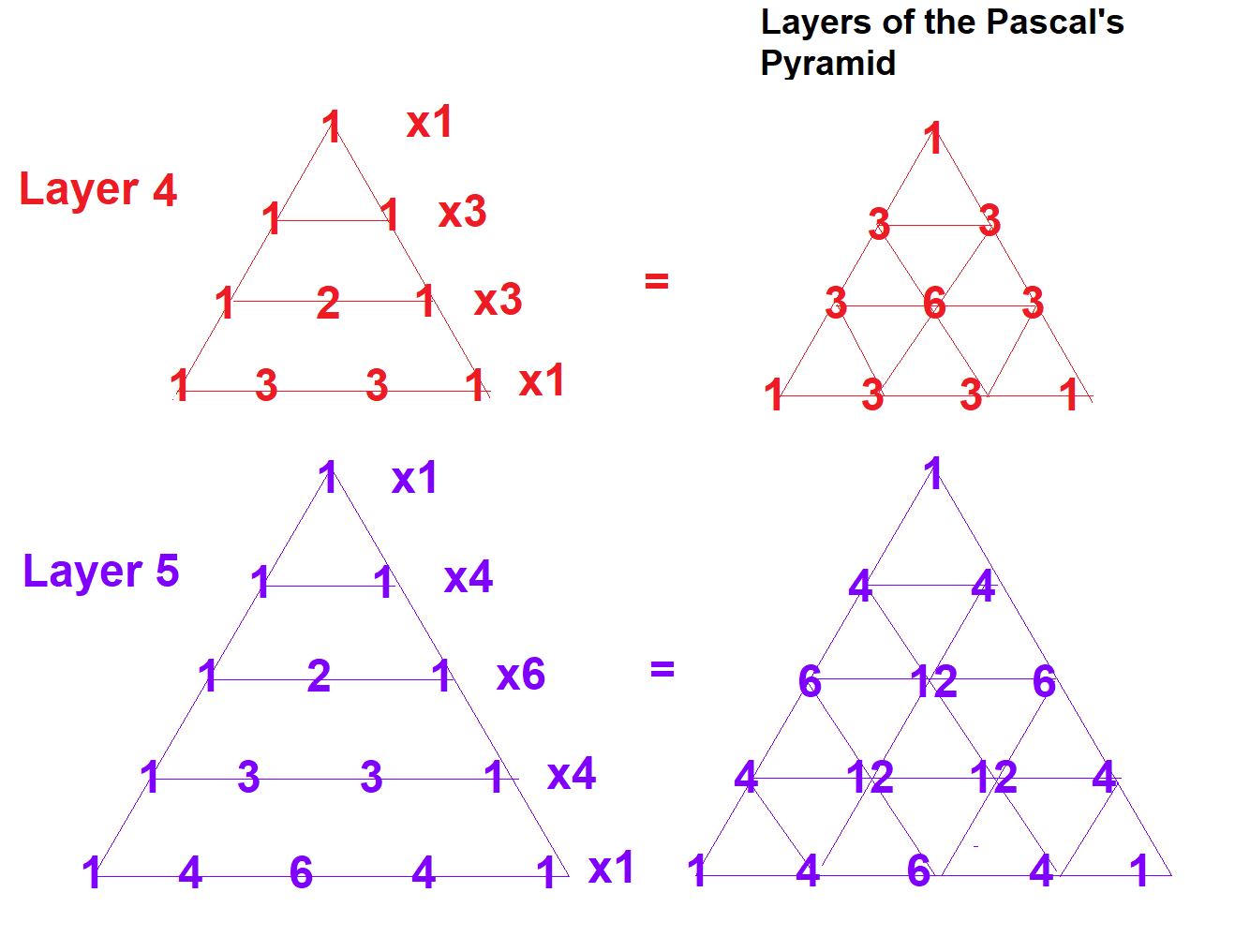}
  \caption{$4$-th and $5$th layers.}
\end{subfigure}
\caption{Construction of the first $5$ layers of the Pascal's pyramid.}
\label{fig:fig3}
\end{figure}

The Pascal's pyramid is originally constructed by Staib \cite{Staib1978} as a $3$-dimensional generalisation of the Pascal's triangle. It can be constructed as follows. For the $n$-th layer in the Pascal's pyramid, take the first $n$ rows of the Pascal's triangle and multiply each row with elements from the last row. Figure \ref{fig:fig3} shows the construction of the first five layers.

\begin{figure}[H]
\begin{subfigure}{.5\textwidth}
  \centering
  \includegraphics[width=0.9\linewidth]{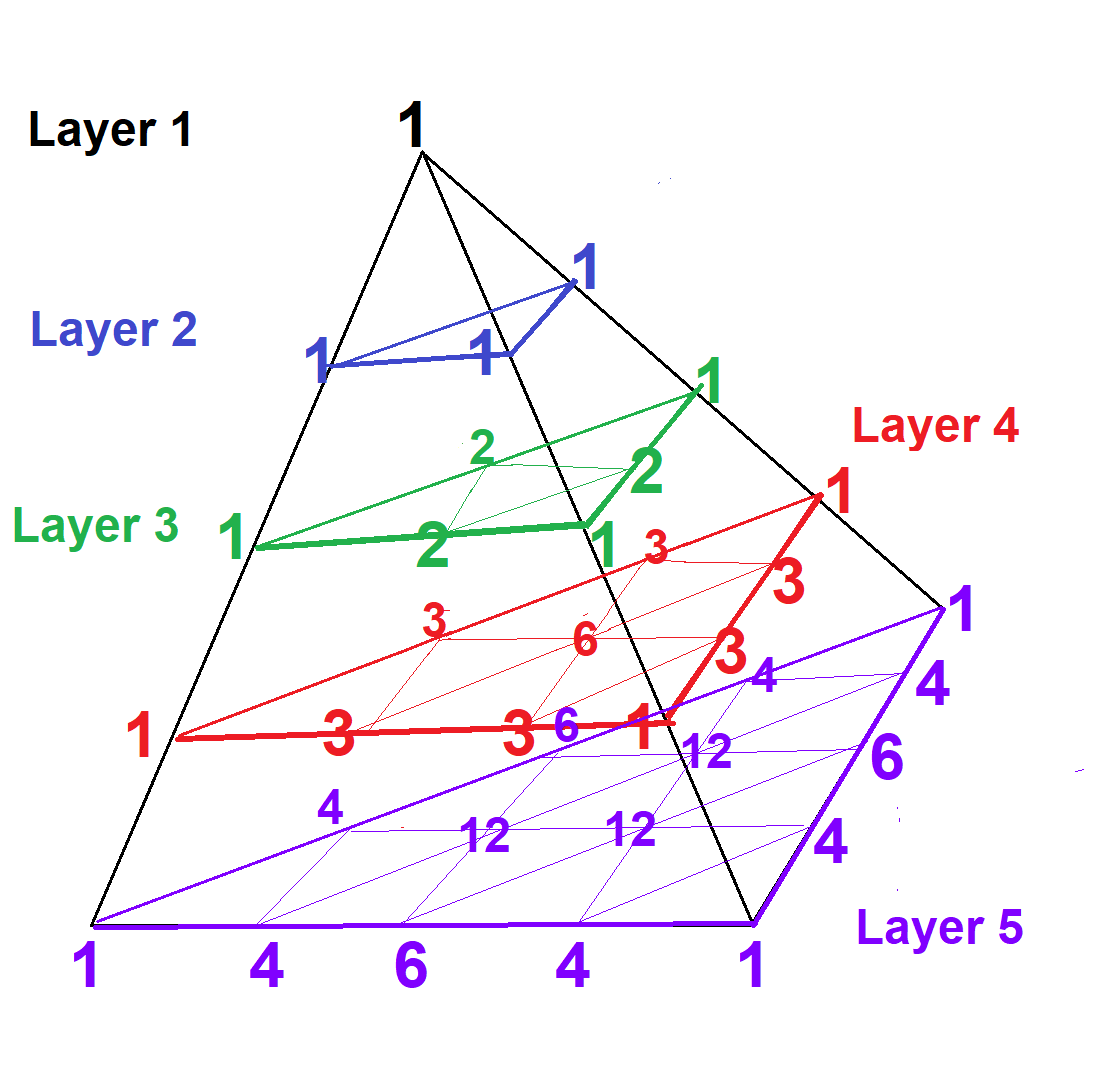}
  \caption{First $5$ layers of the Pascal's pyramid.}
 \label{Pascal pyramid}
\end{subfigure}
\begin{subfigure}{.5\textwidth}
  \centering
  \includegraphics[width=1.0\linewidth]{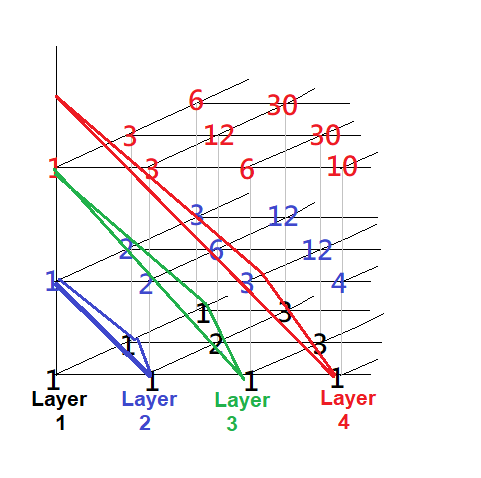}
  \caption{$4$-th and $5$th layers.}
  \label{pascal pyramid to cube}
\end{subfigure}
\caption{Pascal's pyramid}
\label{fig:pascal pyramid}
\end{figure}

Figure \ref{Pascal pyramid} presents the first 5 layers of the Pascal's pyramid. 
 
 Properties of the Pascal's pyramid include
 \begin{itemize}
 \item There is a three way symmetry in each layer.
 \item Sum of all numbers in the $n$-th layer is $3^{n-1}$. 
 \end{itemize}
 
 We can map layers of the Pascal's pyramid to the Pascal's cube. Mapping of the first three layers are shown in Figure \ref{pascal pyramid to cube}. Each layer of the Pascal's pyramid becomes a diagonal cross-section in the Pascal's cube. 

\begin{figure}[H]
\begin{subfigure}{.5\textwidth}
  \centering
  \includegraphics[width=0.9\linewidth]{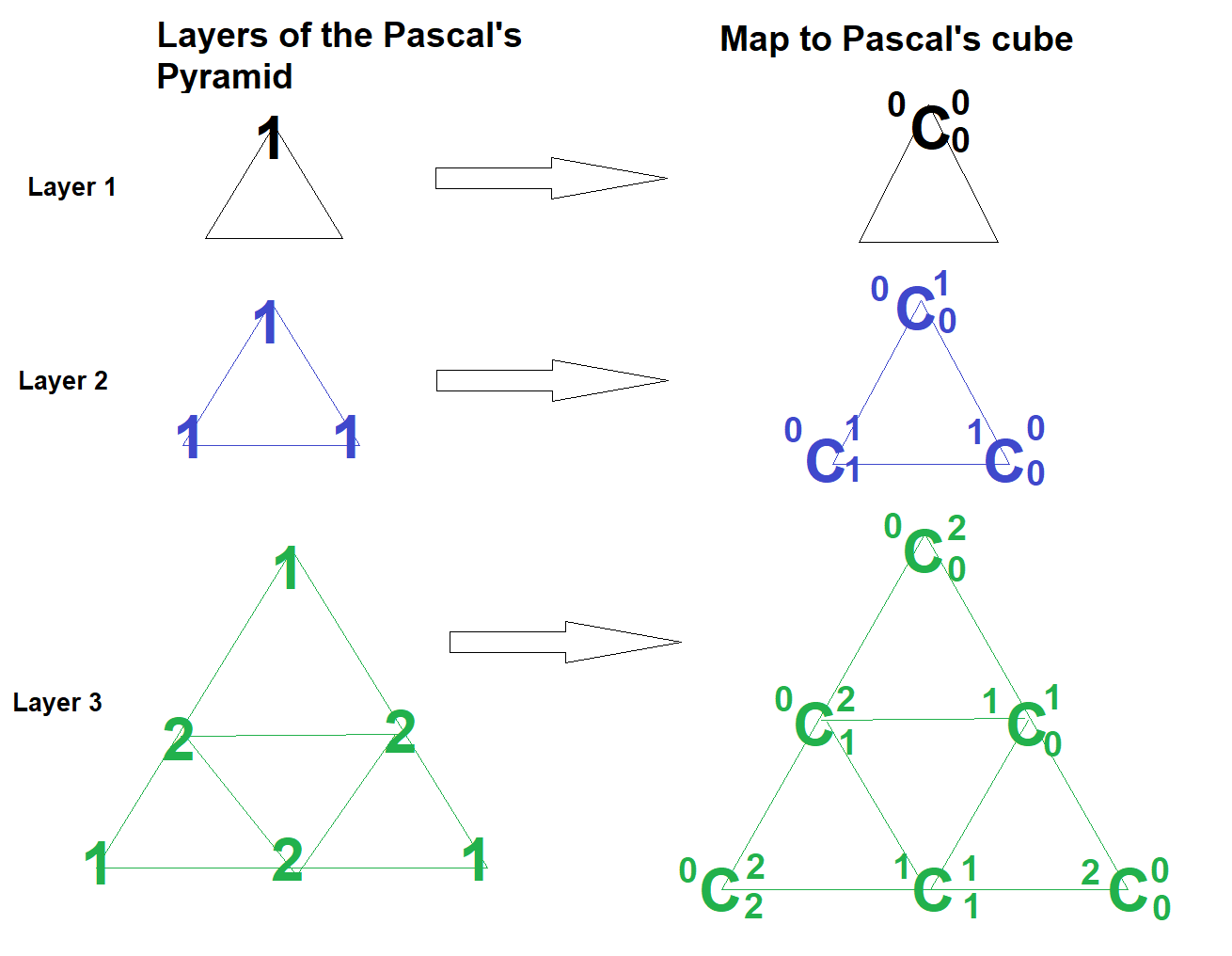}
  \caption{$1$st to $3$rd layer.}
 \label{1 to 3 com}
\end{subfigure}
\begin{subfigure}{.5\textwidth}
  \centering
  \includegraphics[width=1.0\linewidth]{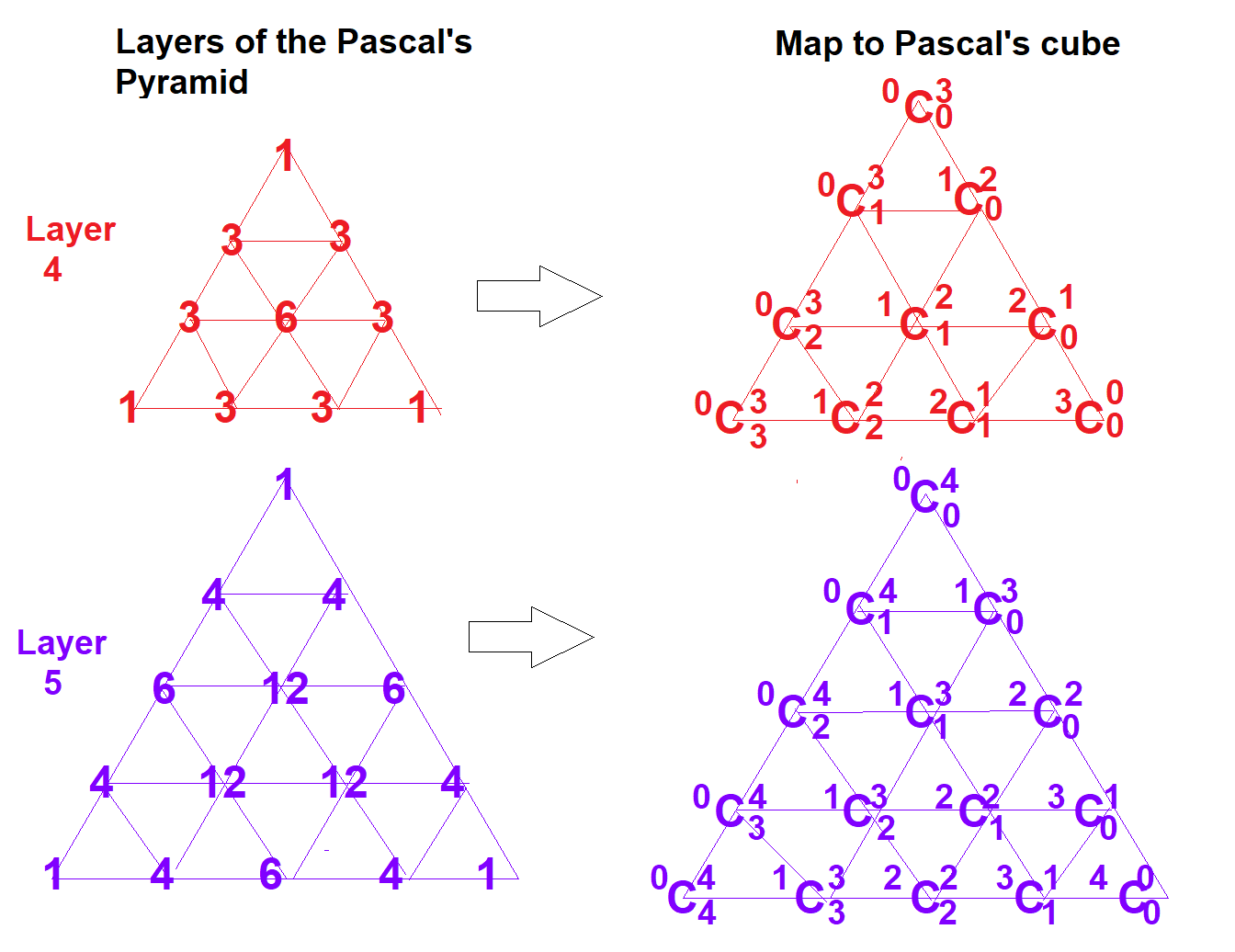}
  \caption{$4$th and $5$th layers.}
  \label{4 5 com}
\end{subfigure}
\caption{Mapping generalised binomials to the first $5$ layers of the Pascal's pyramid.}
\label{fig:pascal pyramid com}
\end{figure}

We now map the layers of the Pascal's pyramid to the Pascal's cube and represent elements of the Pascal's pyramid using generalised binomials. The first $5$ layers are shown in Fig \ref{fig:pascal pyramid com}.

\begin{theorem}
By comparing elements in the Pascal's pyramid with extended binomial coefficients in the Pascal's cube, we can re-define the extended binomial coefficients using the following relationship:
\begin{equation}\label{extended binomial 2}
{}^{z}C^{x}_{y} = C^{y+z}_{z}\times C^{x+z}_{y+z}.
\end{equation} 
\end{theorem}

Eqn. \ref{extended binomial 2} is equivalent to the equation in Theorem 2.1, however Eqn. \ref{extended binomial 2} is much easier to compute with. 

\section{Probability of a random walk on the Pascal's pyramid}

\begin{figure}[H]
\begin{subfigure}{.5\textwidth}
  \centering
  \includegraphics[width=0.9\linewidth]{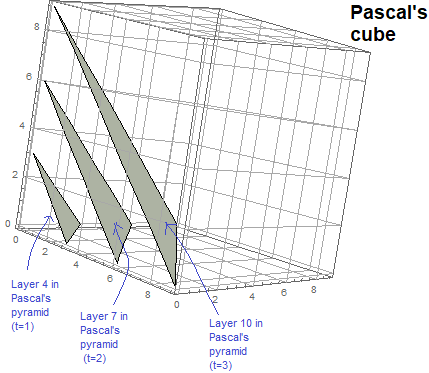}
  \caption{Layers of the Pascal's pyramid.}
 \label{cross section}
\end{subfigure}
\begin{subfigure}{.5\textwidth}
  \centering
  \includegraphics[width=1.0\linewidth]{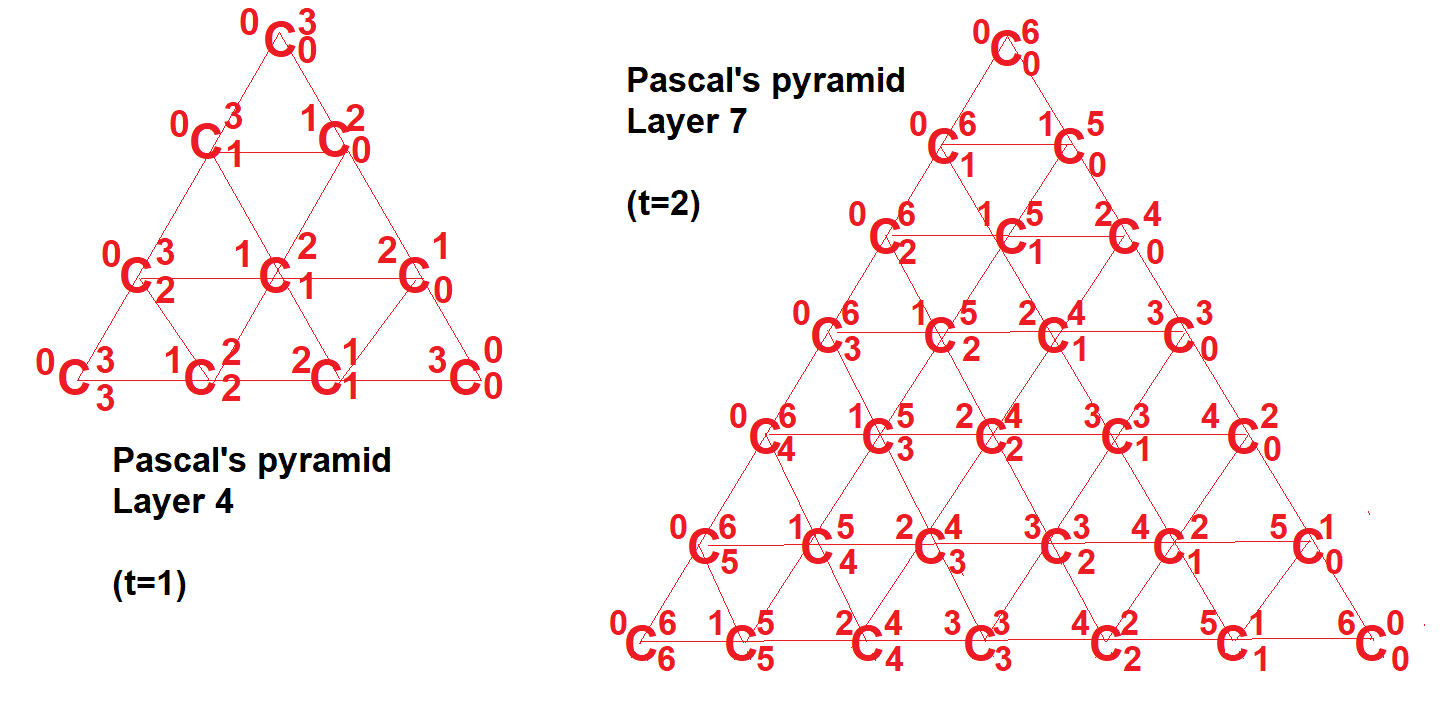}
  \caption{Layer 4 and layer 7 of the Pascal's pyramid.}
  \label{4 7 layer}
\end{subfigure}
\caption{Pascal's pyramid.}
\label{fig:pascal pyramid}
\end{figure}

Let us investigate the dispersion of a particle using the Pascal's pyramid. Consider the convention that at time $t$, the probability distribution of the particle depends on the extended binomial coefficients on the $3t+1$-th layer of the Pascal's pyramid, as shown in Fig. \ref{cross section}. Fig. \ref{4 7 layer} shows the layers of Pascal's pyramid corresponding to $t=2$ and $t=3$ in terms of extended binomial coefficients. 

\begin{figure}[H]
  \includegraphics[width=\linewidth]{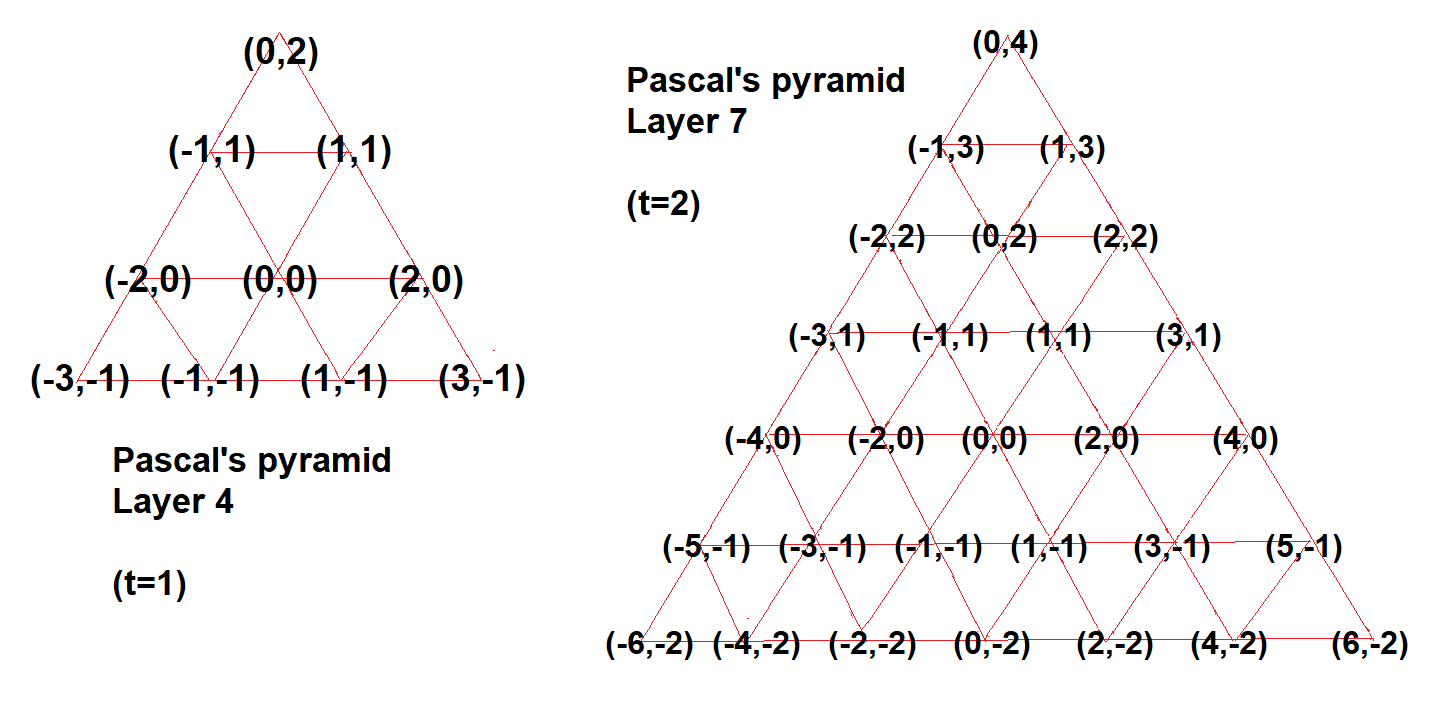}
  \caption{$(x,y)$ coordinate on layer $4$ and layer $7$ of the Pascal's Pyramid.}
  \label{fig:Pascal pyramid 4 and 7 coordinate}
\end{figure}

We construct a coordinate system on layers of the Pascal's pyramid where $x$ corresponds to the rows while $y$ corresponds to the columns. Fig. \ref{fig:Pascal pyramid 4 and 7 coordinate} demonstrates coordinate $(x,y)$ on the $4$-th and the $7$-th layers of the Pascal's pyramid, where the extended binomial coefficient at $(0,0)$ is ${}^{t}C^{2t}_{t}$.   

The extended binomial coefficient at $(x,y)$ on the $3t+1$-th layer of the Pascal's pyramid is given by
\[ {}^{t+\frac{1}{2}(x-y)}C^{2t-\frac{1}{2}(x-y)}_{y+t}.\]

We define a probability function $P(x,y)$ on the $3t+1$-th layer as 
\begin{equation}
P(x,y)=\frac{{}^{t+\frac{1}{2}(x-y)}C^{2t-\frac{1}{2}(x-y)}_{y+t}}{3^{3t}},
\end{equation} 
where the scaling factor $3^{3t}$ is the sum of all entries in the $3t+1$-th layer of the Pascal's pyramid.

%
%
%

\begin{figure}[H]
\begin{subfigure}{.5\textwidth}
  \centering
  \includegraphics[width=1.0\linewidth]{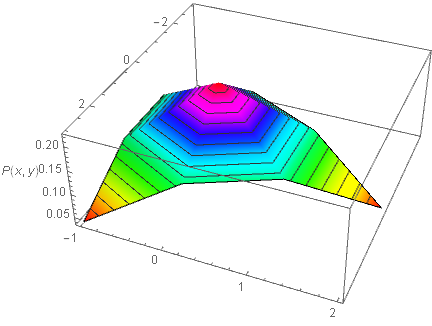}
  \caption{Probability distribution at $t=1$.}
 \label{P1}
\end{subfigure}
\begin{subfigure}{.5\textwidth}
  \centering
  \includegraphics[width=1.0\linewidth]{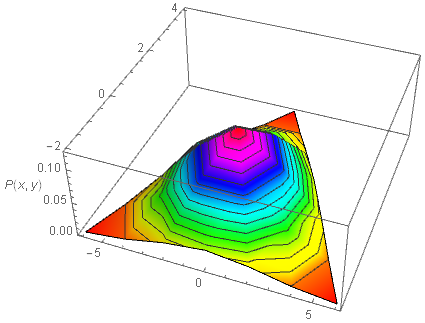}
  \caption{Probability distribution at $t=2$.}
  \label{P2}
\end{subfigure}
\begin{subfigure}{.5\textwidth}
  \centering
  \includegraphics[width=1.0\linewidth]{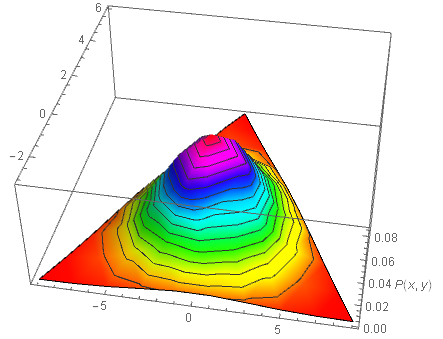}
  \caption{Probability distribution at $t=3$.}
  \label{P3}
\end{subfigure}
\caption{Probability distribution of a particle from $t=1$ to $t=3$.}
\label{fig:Probability distribution}
\end{figure}

Fig. \ref{fig:Probability distribution} shows $P(x,y)$, the probability distribution of the particle at $t=1$, $t=2$ and $t=3$. 

\section{Heat equation}

\begin{figure}[H]
  \includegraphics[width=\linewidth]{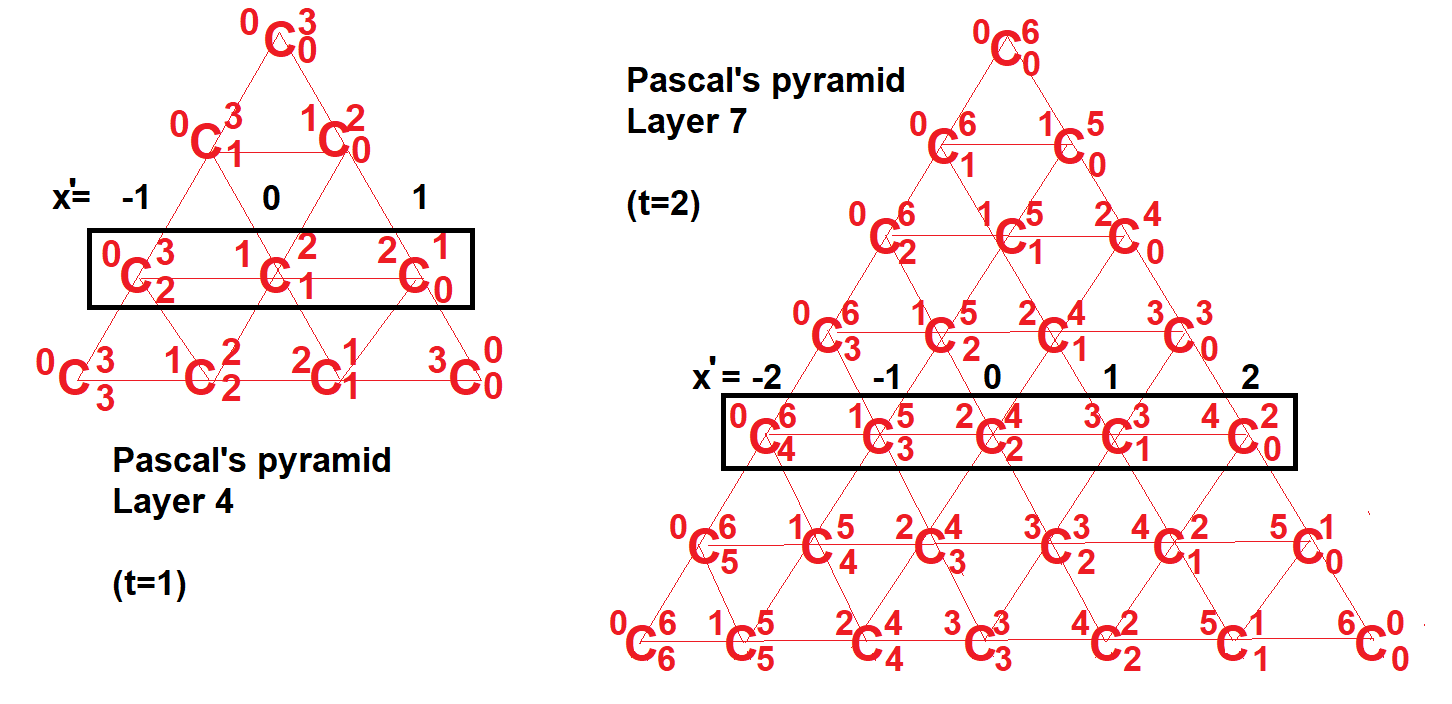}
  \caption{Layer 4 and layer 7 of the Pascal's pyramid.}
  \label{fig:layert3and7}
\end{figure}

Let us define a stochastic process $P(x',t)$ on the Pascal's pyramid, with $t$ corresponds to the $3t+1$-th layer of the Pascal's pyramid, and $x'$ corresponds to the position in the middle row (i.e. $y=0$ and $x'=x/2$) as shown in Fig. \ref{fig:layert3and7}. The extended binomial coefficient corresponds to coordinate $x'$ at time $t$ is given by ${}^{t+x'}C^{2t-x'}_{t}$.

\begin{definition}
Let $P(x',t)$ be a discrete-time process defined by
\begin{equation}
P(x',t) = \frac{{}^{t+x'}C^{2t-x'}_{t}}{3^{3t}},
\end{equation}
where $-t\le x \le t$, and $x'\in \mathbb{Z}$. 
\end{definition}

\begin{figure}[H]
  \includegraphics[width=0.8\linewidth]{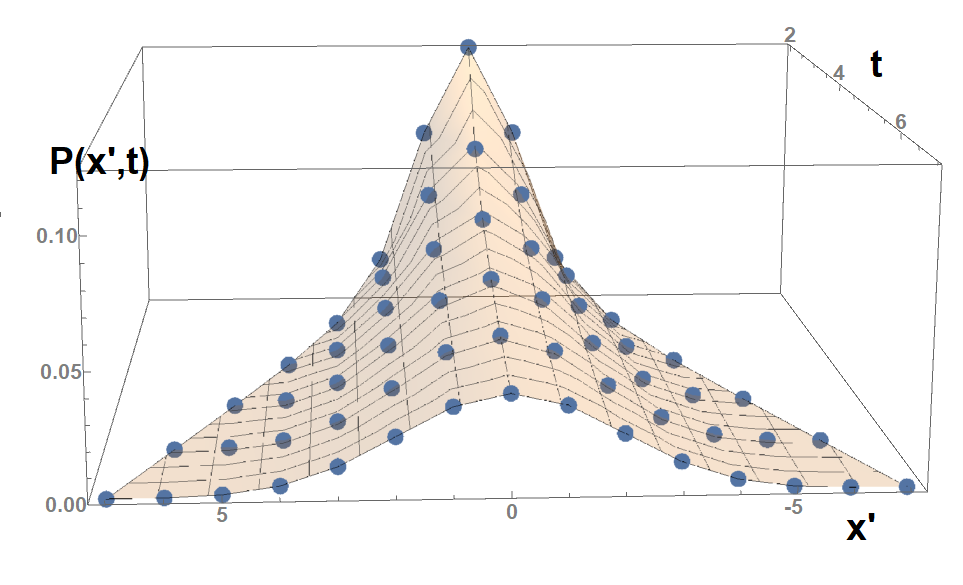}
  \caption{Spreading of $P(x',t)$ in the space-time lattice.}
  \label{fig:probability distribution}
\end{figure}

Fig \ref{fig:probability distribution} demonstrates the evolution of $P(x',t)$ from $t=2$ to $t=7$.

\begin{theorem}
For large $t$ and $t \gg x'$, $P(x',t)$ satisfies the following diffusion equation:
\begin{equation}\label{dispersion eqn}
\frac{\partial P(x',t)}{\partial t}=\frac{1}{2}\frac{\partial^{2}P(x',t)}{\partial x'^{2}},
\end{equation}
\end{theorem}

\begin{proof}
First of all,
\begin{equation*}
P(x',t) = \frac{{}^{t+x'}C^{2t-x'}_{t}}{3^{3t}} =  \frac{C^{2t+x'}_{t+x'} \times C^{2t-x'}_{t}}{3^{3t}}=\frac{3t!}{(t+x')! \times t! \times (t-x')! \times 3^{3t}}
\end{equation*}
$P(x',t)$ is a discrete function, here we replace $\Gamma(n+1)$ for $n!$ in order to differentiate $P(x',t)$:
\begin{equation*}
P(x',t)=\frac{\Gamma(3t+1)}{\Gamma(t+x'+1)\times\Gamma(t+1)\times\Gamma(t-x'+1)\times 3^{3t}}.
\end{equation*}
Assuming $t$ is very large and $t\gg x'$, then
\begin{equation*}
\frac{\partial P(x',t)}{\partial t}=P(x',t)\times \left( -\frac{1}{2(t+x')} - \frac{1}{2(t-x')} \right), 
\end{equation*}
and
\begin{equation*}
\frac{\partial^{2}P(x',t)}{\partial x'^{2}} = P(x',t) \times \left( -\frac{1}{t+x'} -\frac{1}{t-x'}\right).
\end{equation*}
 Thus $P(x',t)$ satisfies Eqn. \ref{dispersion eqn}, a heat equation. 
\end{proof}

\section{Conclusion}

In this paper, we presented a computationally useful way to explore Pascal's pyramid, a $3$ dimensional generalisation of the Pascal's triangle. We constructed a stochastic process that resembles probability distribution of a particle propagating on the Pascal's pyramid, and showed that with a certain constraint, the stochastic process satisfies the heat equation. 

It is well-known that Schr\"{o}dinger equation can be derived from the heat equation when the time becomes imaginary. However it is unclear if the approach presented here can be applied to the Schr\"{o}dinger equation. In this paper we proposed a possible way to connect stochastic processes via the Pascal's pyramid to quantum mechanics.

\section*{Acknowledgement}

I would like to thank my parents for their love and support. I would also like to thank David Ridout for kindly endorsing me on arXiv so I can submit this paper. This is a little project that I do for fun after work. Most of the graphs are completed thanks to Mathematica, it is also a great fun to revise and explore the program. It is a simple project and the mathematics has probably already been explored and presented in another paper that I am not aware of. Please feel free to contact me if this is the case or to give me feedback.

%






\bibliographystyle{IEEEtranN}



\end{document}